\documentclass{aastex}
\usepackage{spr-astr-addons}             

\RequirePackage{color}                   

\begin{document}

\title{Photometric Calibration of the $[\alpha/$Fe] Element:\\
 II. Calibration with SDSS Photometry}
\slugcomment{Not to appear in Nonlearned J., 45.}
\shorttitle{Calibration with SDSS Photometry}
\shortauthors{E. Yaz G\"ok\c ce, S. Bilir, S. Karaali, O. Plevne}

\author{E. Yaz G\"ok\c ce \altaffilmark{1}}
\altaffiltext{1}{Istanbul University, Faculty of Science, Department 
of Astronomy and Space Sciences, 34119 University, Istanbul, Turkey\\
\email{esmayaz@istanbul.edu.tr}}

\author{S. Bilir \altaffilmark{1}}
\altaffiltext{1}{Istanbul University, Faculty of Science, Department 
of Astronomy and Space Sciences, 34119 University, Istanbul, Turkey\\}

\author{S. Karaali \altaffilmark{1}} 
\altaffiltext{1}{Istanbul University, Faculty of Science, Department 
of Astronomy and Space Sciences, 34119 University, Istanbul, Turkey\\}

\and
\author{O. Plevne \altaffilmark{2}}
\altaffiltext{1}{Istanbul University, Department of Astronomy and Space Sciences, Graduate School of Science and Engineering, Istanbul University, 34116, Beyaz{\i}t, Istanbul, Turkey\\}

\begin{abstract}  
We present the calibration of the [$\alpha/$Fe] element in terms of ultra-violet excess for 465 dwarf stars with spectral type F0-K2. We used a single calibration, fitted to a third degree polynomial with a square of the correlation coefficient 0.74 and standard deviation 0.05 mag, for all stars due to their small colour range, $0.1<(g-r)_0\leq 0.6$ mag, and high frequency in the blueward of the spectrum which minimize the guillotine effect. Our calibration provides [$\alpha/$Fe] elements in the range $(-0.05, 0.35]$ dex. We applied the procedure to a high-latitude field, $85^\circ \leq b \leq 90^\circ$ with size 78 deg$^2$ and we could estimate the [$\alpha/$Fe] elements of 23,414 dwarf stars which occupy a Galactic region up to a vertical distance of $z=9$ kpc. We could detect a small positive gradient, $d[\alpha/{\rm Fe}]/dz=+0.032 \pm0.002$ dex kpc$^{-1}$, for the range $0<z<5$ kpc, while the distribution of the [$\alpha/$Fe] element is flat for further $z$ distances.
\end{abstract}


\section{Introduction}
Metallicity of a star is an important parameter for understanding the formation and evolution of different components of our Galaxy. The abundance of the iron element relative to the abundance of the hydrogen, [Fe/H], has been widely used for a long time for separation of the stars into two distinct populations, i.e. an iron-rich disc and an iron-poor halo populations. The need of a third component for fitting of the space densities of the Galaxy \citep{Gilmore83} revealed a third population with intermediate iron abundance. The source of the iron-peak elements is mainly the Type Ia supernovae which are produced at large timescales, a few Gyr. There is another set of elements, i.e. alpha ($\alpha$) elements, ($^{4}$He-nuclei). Most of these elements are produced from the Type II supernovae in a short timescales, 20 Myr \citep{Wyse88}. However, there are some contributions from Type Ia supernovae. The combination of the [$\alpha/$Fe] element with the [Fe/H] one reveals an interesting picture, i.e. [$\alpha/$Fe] element has a flat distribution for metal-poor and old stars and it has a negative gradient for relatively metal-rich stars. 

The metallicity of a star can be determined either spectroscopically or photometrically. Nearby dwarfs and distant giants have the advantage of high-resolution spectra. Thanks to \cite{Roman55} who interpreted the weakness of the metallicity lines as the metallicity determination of a star, i.e. the metallicity of a star can be measured by its ultra-violet ({\it UV})-excess, $\delta(U-B)$. This procedure can be applied also to distant stars \citep[e.g.][]{Karaali03a, Karaali03b, Karaali03c, Karaali05, Guctekin16}. Metallicity determination by using the second procedure has been carried out by many researchers \citep[cf.][and references therein]{Karaali16}. 

The alpha elements became important phenomena since about the beginning of this century. One can find a large set of various alpha elements in the Hypatia catalogue \citep{Hinkel14}. This catalogue is a collection of many elements published in 69 studies. The least number of elements measured in these studies is two \citep{Ecuvillon06,Caffau11}, while the biggest one is 33 \citep{Galeev04}. The numbers of stars observed in different studies are also different, i.e. \cite{Neuforge97} and \cite{Porto08} measured only two stars, while \cite{Petigura11} and \cite{Valenti05} derived as much as 914 and 1002 stars, respectively. The total number of stars in the Hypatia catalogue is 8821. However, there are many overlapping stars in different studies.

There are four more catalogues published in the following studies which cover a large number of stars with alpha elements: \citet[][hereafter V04]{Venn04}: 780 stars, \citet[][hereafter B14]{Bensby14}: 714 stars, \citet[][hereafter R06]{Reddy06}: 176 stars, and \citet[][hereafter N10]{Nissen10}: 100 stars. The study of \citet[][hereafter S02]{Stephens02} consists of 56 halo stars with eight alpha elements. Also, there are many overlapping stars in these catalogues.   

The maximum value of the alpha elements in the cited studies is [$\alpha/$Fe] $\sim$ 0.4 dex. However, \cite{Jackson14} showed that the metal-poor halo stars in the Galactic halo as observed by the {\it Gaia}-ESO survey may be as rich as [$\alpha/$Fe]=0.6 dex. Observation of the alpha elements has also been carried out in other recent large surveys, i.e. RAVE \citep{Steinmetz06}, APOGEE \citep{Wilson10}, GALAH \citep{Heijmans12}, and GIBS \citep{Zoccali14}.

The pioneers of the alpha element observers used their data to separate the Galactic stars into thin disc, thick disc, and halo populations (V04, B14, R06, N10). However, in recent years, the alpha elements have been the issue of a different subject i.e. they are used to confirm the radial migration simulations and disc flaring \citep[cf.][]{Minchev17}.

As noted in the foregoing paragraph, the alpha measurements are limited with distance. The first time, we could calibrate the alpha elements -combination of [Mg/Fe], [Ca/Fe], [Ti/Fe] and [Na/Fe]- in terms of $UV$-excess, $\delta(U-B)$ which can also be applied to distant stars \citep[][hereafter K16]{Karaali16}. In K16, we estimated synthetic alpha elements, $[\alpha$/Fe]$_{syn}$, and {\it UV}-excesses, $\delta_{syn}$, using the Dartmooth Stellar Evolution Database \citep{Dotter08} and compared their distribution with the one estimated by our calibration. The agreement between two mentioned distributions encouraged us to obtain a similar calibration with the $ugr$ data. We aim to calibrate the same alpha elements in terms of $\delta(u-g)$, the $UV$-excess defined with $ugr$ photometry, and apply it to a large set of F and G spectral type stars. We organized the paper as follows: The data are given in Section 2 and the procedure is explained in Section 3. The results are presented in Section 4 and finally, Section 5 is devoted to a summary and discussion.

\section{Data}
The sample star and their [$\alpha/$Fe] elements are adopted from K16 as explained in the following. The catalogues of V04, B14, R06 and N10 cover a large number of stars with different alpha-elements, i.e. 780, 714, 176 and 100 stars, respectively. The catalogue of V04 is a collection of 15 studies, also there is a large overlapping of the stars in the four catalogues. K16 gave priority to the stars in V04 and B14 and applied a series of constraints to obtain the final set of data available for alpha element calibration, i.e. they reduced the multiplicity of the stars to a single one, they considered only the stars for which the alpha elements [Mg/Fe], [Ca/Fe], [Ti/Fe], and [Na/Fe] are available, and they omitted the giants and the dwarfs without $U-B$ and $B-V$ colour indices. Thus, their sample reduced to 589 stars (their Table 1). They de-reddened the colours $U-B$ and $B-V$ and plotted the [$\alpha/$Fe] elements versus the {\it UV}-excesses $\delta$ of the stars, separated into nine sub-samples, and they rejected the stars which show large scatter in the $[\alpha$/Fe$]-\delta$ diagram. These stars are candidates for binarity, variable, double or multiple and chromospheric active stars. These constraints reduced the number of stars to 541. K16 calibrated the [$\alpha/$Fe] elements in terms of $\delta$ and reproduced the alpha elements, $[\alpha/{\rm Fe}]_{rep}$, in the reduced sample by replacing their {\it UV}-excesses into this calibration. Finally, they omitted stars with (original) [$\alpha/$Fe] elements which lie out of the interval $[\alpha/{\rm Fe}]_{rep}\pm \Delta[\alpha/\rm {Fe}]$ where $\Delta[\alpha/{\rm Fe}]$ corresponds to the mean of the residuals, the differences between the original and reproduced alpha elements, larger than the standard deviation of the residuals. Thus they obtained a sample of 469 dwarfs (their Table 3). This is the sample used in our study. The de-reddened colours, $(U-B)_0$ and $(B-V)_0$, are transformed to the $(g-r)_0$ and $(u-g)_0$ ones by using the following transformation equations derived from the equations of \cite{Chonis08}:

\begin{eqnarray}
(g-r)_0=1.094(B-V)_0-0.248,\\ \nonumber
(u-g)_0=(U-B)_0+0.358(B-V)_0+0.989.
\end{eqnarray}
The {\it UV}-colour indices for the Hyades cluster, $(u-g)_H$, are evaluated by the following fifth degree polynomial calibrated via the $(u-g)_0 \times (g-r)_0$ sequence of the Hyades cluster, where $(u-g)_0$ and $(g-r)_0$ colour indices are transformed from the $(U-B)_0 \times (B-V)_0$ sequence of the Hyades cluster in K16:

\begin{figure}
\begin{center}
\includegraphics[scale=0.5, angle=0]{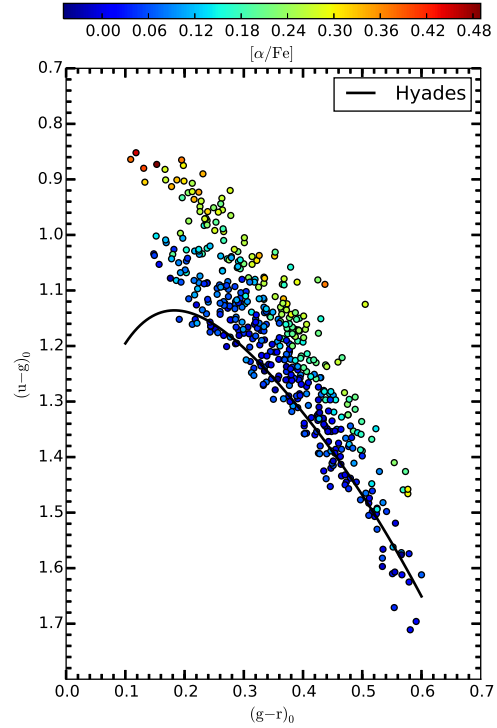}
\caption[]{$(u-g)_{0} \times (g-r)_{0}$ two-colour diagram for the sample stars as a function of the [$\alpha/$Fe] abundances (filled circles) and the Hyades fiducial sequence (solid curve). Colour scale of [$\alpha/$Fe] abundances is given at the top of the figure.} 
\end{center}
\end {figure}

\begin{figure}
\begin{center}
\includegraphics[scale=0.4, angle=0]{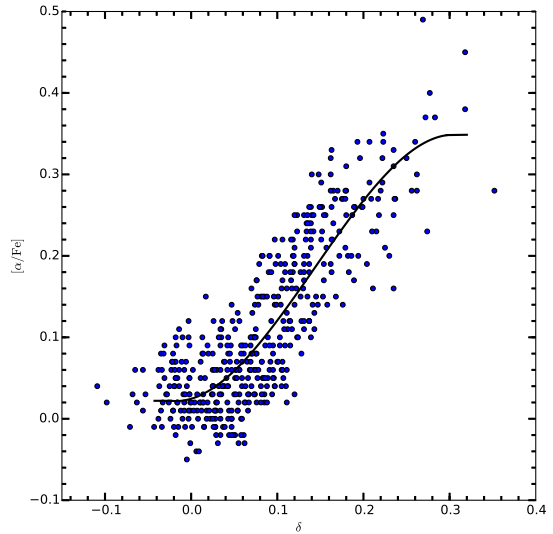}
\caption[]{Calibration of the [$\alpha/$Fe] in terms of $\delta$.} 
\end{center}
\end {figure}

\begin{eqnarray}
(u-g)_{H}=-23.117 X^{5}+56.116 X^{4} -53.052 X^{3}\\ \nonumber
+26.039 X^{2}-5.4432 X+1.526.
\end{eqnarray}
where $X=(g-r)_0$. The $(u-g)_0 \times (g-r)_0$ two-colour diagram for the sample stars and the fiducial sequence of the Hyades cluster are given in Fig. 1. Possible errors in the adopted colour transformation equations are estimated as follows. We iterated the differences between $g-r$, and $u-g$, colours estimated with and without the errors in \cite{Chonis08}, 100,000 times, and evaluated the following means, standard deviations and standard errors: $\langle \Delta (g-r)\rangle=0$, $\sigma_{(g-r)}=0.02$, $\sigma_{err}=0$ mag; $\langle (u-g)\rangle=0$, $\sigma_{(u-g)}=0.12$, $\sigma_{err}=0$ mag. The transformation equations between the SDSS and {\it UBV} systems used in our study are not metallicity dependent. However, we do not expect any systematic error in the $u-g$ and $g-r$ colours as explained in the following: it is the $u$ band which measures the metallicity of a star photometrically, and as stated by \cite{Chonis08}, the passband of the $u$ filter agrees most closely to the Johnson-Cousins' passband; $U=u-0.854\pm0.007$. The [$\alpha/$Fe] elements are the mean values of the [Mg/Fe], [Ca/Fe], [Ti/Fe] and [Na/Fe] elements which are taken from the studies cited in this section. The {\it UV}-excess of a star, $\delta(u-g)=(u-g)_H-(u-g)_*$, is evaluated in this study where $(u-g)_*$ and $(u-g)_H$ are the {\it UV}-colours of a given star and the Hyades one with the same $(g-r)_0$ colour. We limited the $(g-r)_0$ colour range with $0.1<(g-r)_0\leq 0.6$ mag corresponding to a spectral type range F0-K2. Hence the number of the sample stars reduced to 465 whose distribution into the studies mentioned above is as follows: V04 (227), B14 (196), R06 (25) and N10 (17). The ranges of the $(u-g)_0$, $\delta(u-g)$ and [$\alpha/$Fe] are $0.85<(u-g)_0<1.71$, $-0.11<\delta(u-g)<0.30$ mag and $-0.05<[\alpha/{\rm Fe}]<0.35$ dex, respectively. 

\section{The Procedure}
\subsection{Calibration of [$\alpha/$Fe] Element in terms of $\delta(u-g)$}

We adopted the procedure in K16 for our calibration with a small modification, however, i.e. we used a single calibration for all stars instead of nine calibrations used in K16. The reason of this modification is that the colour range of the sample stars is relatively small, $0.1<(g-r)_0\leq0.6$ mag. Additionally, the majority of the sample stars have bluer colours (Fig. 1) where the guillotine effect is rather small. As stated in \cite{Wildey62}, \cite{Sandage69}, \cite{Carney79}, \citet{Karaali03a, Karaali05, Karaali11}, and \citet{Guctekin16}, the stars which are mostly affected by the guillotine effect are the red ones which is not the case in our work. We confirm our argument by referring the numerical values of the correction factors in \cite{Sandage69} which reduce the observed {\it UV}-excesses to the one for the colour $(B-V)_0=0.60$ mag. The range of this factor which covers the interval $0.35\leq(B-V)_0 \leq 1.10$ is between 1.00 and 2.58. However, our star sample is limited with $0.1<(g-r)_0 \leq0.6$ mag or $0.35<(B-V)_0<0.77$. Hence, the corresponding range of the correction factors is between 1.00 and 1.24, and even narrower for 87$\%$ of the sample stars, i.e. 1.00-1.15. Then, if we adopt the maximum value, 1.15, for the correction factor and apply it to a star with biggest observed {\it UV} value, $\delta=0.30$ mag, we obtain the value 0.34 whose difference from the observed value (0.04 mag) is within the photometric errors. 

We should emphasize that a single calibration covers larger number of stars relative to a series of calibrations for the same sample stars which increases the quality of the calibration. Thus, we plotted the [$\alpha/$Fe] elements of all the sample stars versus $\delta(u-g)$, the difference between the $(u-g)_0$ colours of a given star and the Hyades star with the same $(g-r)_0$ colour (Fig. 2) and adopted a third degree polynomial for their calibration as in the following (hereafter we will use the symbol $\delta$ for the {\it UV}-excess): 
\begin{eqnarray}
[\alpha/Fe]=-19.608\delta^3+8.436\delta^2+0.314\delta+0.025.
\end{eqnarray}
The squared of the correlation coefficient of our calibration and the standard deviation are $R^2=0.740$ and $\sigma=0.051$ dex which promise accurate estimation for the [$\alpha/$Fe] element via {\it UV}-excess $\delta$. The ranges of $\delta$ and [$\alpha/$Fe] for the calibration curve are $-0.11\leq \delta \leq 0.30$ mag and $-0.05\leq [\alpha$/Fe]$\leq 0.35$ dex. 

\subsection{Application of the Procedure}
\subsubsection{The star sample}
The star sample consists of 23,414 F and G dwarfs taken from \cite{Guctekin17} which were provided by a series of constraints explained in the following. The star sample was defined with Galactic coordinates $85^\circ \leq b \leq 90^\circ$, $0^\circ \leq l \leq 360^\circ$ and size 78 deg$^2$. There are 1,973,575 objects with de-reddened $ugriz$ magnitudes in the survey DR 12 of SDSS III \citep{Alam15} within this field. The data were restricted with $g_0 \leq 23$ and $(u-g)_0>0.5$ mag to exclude the extra–galactic objects. Then, the following equation of \cite{Juric08} was adopted to reject the stars which were large scattered in the $(g-r)_0 \times (r-i)_0$, i.e. stars beyond $\pm2\sigma$ of this equation:

\begin{eqnarray}
(g-r)_0 =1.39(1-\exp[-4.90(r-i)_0^3-2.45(r-i)_0^2\\ \nonumber
-1.68(r-i)_0-0.05].
\end{eqnarray}
This is carried out by iso-density closed curves which cover larger number of stars as one moves outward of the curve with Eq. (4). The iso-density curve which covers 411,512 stars is adopted as the limit between our sample stars and the rejected ones. Actually, the ratio of 411,512 to the total number of stars, 433,636 in Fig. 10 of \cite{Guctekin17} is 95$\%$ corresponding to the mean $\pm 2\sigma$ in a normal distribution. 

Additionally, 131 giants were identified by using the equations of \citet{Helmi03}, \citep[see also][]{Bilir08} and they were rejected from the sample. A further restriction was applied to the apparent de-reddened magnitude, i.e. $g_0 \leq 20$, to avoid from large errors in magnitude and colours. A final constraint is related to the absolute magnitudes which were estimated via the procedure in the cited paper. We should emphasize that all these constraints were carried out in \cite{Guctekin17}.

\begin{figure}[t]
\begin{center}
\includegraphics[scale=0.55, angle=0]{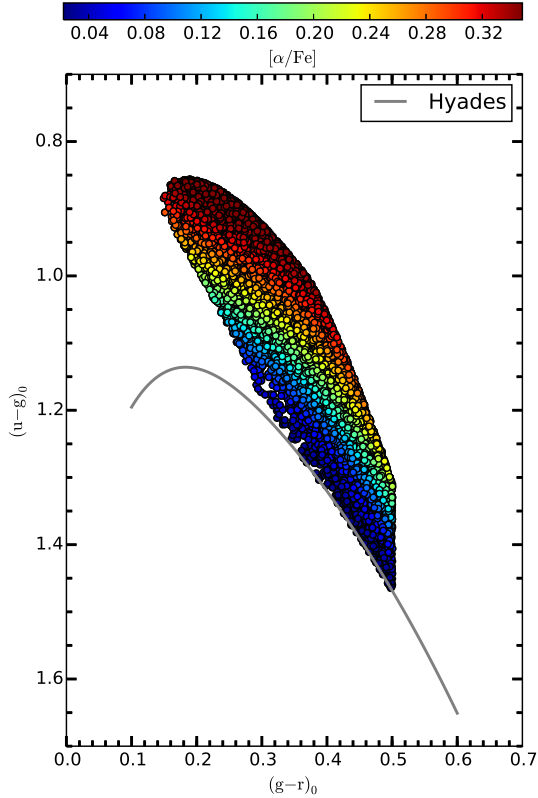}
\caption[]{$(u-g)_0 \times (g-r)_0$ two-colour diagram for the SDSS stars as a function of [$\alpha/$Fe] used in the application of the procedure (filled circles) and the Hyades fiducial sequence (solid curve). Colour scale of the [$\alpha/$Fe] abundances is given at the top of the figure.} 
\end{center}
\end {figure}

\subsubsection{Application of the procedure} 
We applied the procedure to the star sample mentioned in the previous paragraph as explained in the following. We estimated the $UV$-excess, $\delta$, via the procedure given in Section 2 and replaced them in the Eq. (3), thus we obtained their [$\alpha/$Fe] elements. Then, we evaluated their vertical distances relative to the Galactic plane, $z$, as follows. We combined the apparent and absolute magnitudes, $g_0$ and $M_g$, respectively, to evaluate the distances relative to the Sun, $r$, of the sample stars via the Pogson formula and they are reduced to the vertical distances by means of the equation $z=r \times \sin (b)$, where $b$ is the Galactic latitude of the star in question. The two-colour diagram, $(u-g)_0 \times (g-r)_0$, of the stars used in the application of the procedure, is plotted in Fig. 3. 

The distribution of the $[\alpha/{\rm Fe}]$ elements versus $z$ is demonstrated in Fig. 4. One can see a monotonic increasing of the [$\alpha/$Fe] elements with increasing of $z$ up to $\approx 5$ kpc and a flat distribution for larger $z$ distances. Our first interpretation is that the positive gradient, $d[\alpha/{\rm Fe}]/dz=+0.032\pm0.002$ dex kpc$^{-1}$, indicates a transition from the thin disc stars to the thick-disc stars, while the flat one ($-0.005\pm0.002$ dex kpc$^{-1}$) corresponds to the halo stars. 

\begin{figure}[t]
\begin{center}
\includegraphics[scale=0.27, angle=0]{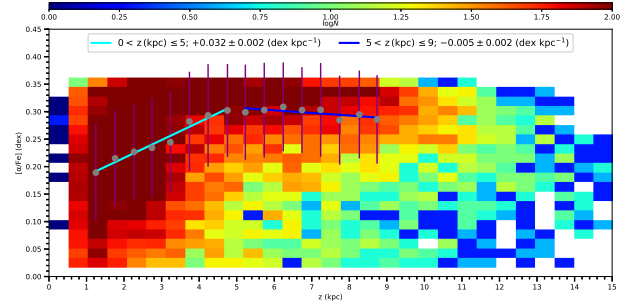}
\caption[]{Colour-coded diagram [$\alpha/$Fe] $\times z$ for the stars used in the application of the procedure. The [$\alpha/$Fe] elements are estimated via the calibration based on the transformation equations.} 
\end{center}
\end {figure}

\subsubsection{Comparison of the spectroscopic and photometric alpha elements}
We used two sets of data and compared the spectroscopic and photometric alpha elements, $[\alpha/{\rm Fe}]_{spec}$ and $[\alpha/{\rm Fe}]_{phot}$, to test the reliability of our calibration. The first set of data consists of the $[\alpha/{\rm Fe}]_{phot}$ elements estimated for 465 calibration stars in this study and the $[\alpha/{\rm Fe}]_{spec}$ elements for 1026 stars taken from the Hypatia catalogue \citep{Hinkel14}. As the $[\alpha/{\rm Fe}]_{phot}$, the $[\alpha/{\rm Fe}]_{spec}$ element is the mean of the [Mg/Fe], [Ca/Fe], [Ti/Fe] and [Na/Fe] elements for a given star. The result is given in Fig. 5, where the $[\alpha/{\rm Fe}]_{spec}$ and $[\alpha/{\rm Fe}]_{phot}$ elements are plotted against spectroscopic iron element, [Fe/H]. The red full circles correspond to the alpha elements taken from the Hypatia catalogue, while the blue ones indicate the alpha elements estimated in our study. The number of metal-poor stars ([Fe/H]$<-1$ dex) in the Hypatia catalogue is about half of a dozen which does not give a chance to us for comparison of the two type of alpha elements. While, for the range ${\rm [Fe/H]} \geq -1$ dex, it seems an agreement between the photometric and spectroscopic alpha elements. One can see a horizontal  feature in the distribution of the alpha elements, corresponding to $[\alpha/{\rm Fe}]_{spec}\cong -0.08$ dex, for the metal-rich stars taken from the Hypatia catalogue. However, such a feature does not effect the agreement just claimed.

\begin{figure}[h]
\begin{center}
\includegraphics[scale=0.55, angle=0]{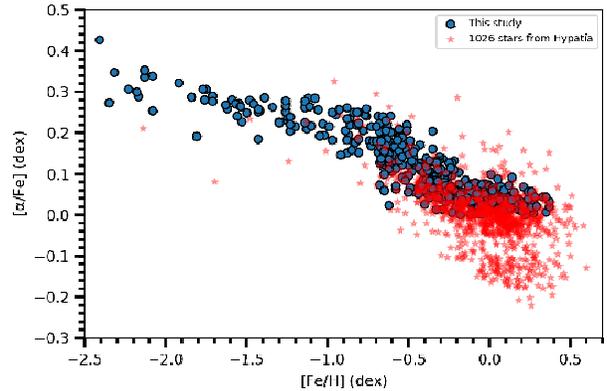}
\caption[]{Comparison of the photometric alpha elements of 465 calibration stars estimated in our study with the spectroscopic ones, for 1026 stars, taken from the Hypatia catalogue, in terms of spectroscopic iron elements.} 
\end{center}
\end {figure}

The second set of alpha elements consists of the Sloan Extension for Galactic Understanding and Exploration data \citep[SEGUE,][]{Yanny09} and those estimated by the calibration (Eq. 3, Fig. 2) in our study. The SEGUE data for 220,851 stars are taken from the SDSS DR12 \citep{Alam15} web page\footnote{http://www.sdss3.org/dr12/data$\_$access/}. We applied a series of constraints, as explained in the following, and obtained a set of stars with alpha elements and $ugr$ magnitudes: The first constraint, $4<\log g <5$, provided us a dwarf sample while the second one, $15<g_0< 17$ mag, rejected the dwarfs with relatively larger errors. The third constraint, $0.1<(g-r)_0<0.6$ mag, is due to adopt the same colour interval used in our calibration. Finally, we used a limitation for the signal to noise ratio, $S/N>50$, for the SEGUE spectra, and reduced the star sample to $N=13,379$. The photometric alpha elements of the sample stars are estimated by adopting their $(g-r)_0$ and $(u-g)_0$ colours to our calibration. We plotted the spectroscopic alpha elements taken from the spectroscopic survey SEGUE and the photometric ones estimated with our calibration in Fig. 6, and compared their distributions to test the reliability of our calibration. However, we should emphasize that the spectra resolution ($R\sim 2000$) in SEGUE is not as high as the ones in other studies, such as the Hypatia catalogue \citep{Hinkel14} or V04, B14, R06 and N10 from which the spectroscopic alpha elements are taken for our calibration.Hence, one can expect small deviations between two distributions. The panel (a) in Fig. 6 shows the distribution of the spectroscopic alpha elements taken from SEGUE versus spectroscopic iron elements. The two contours indicate the number of star within one and two standard deviations, i.e. 68\% and 95\% of the total stars. The distribution of the  photometric alpha elements versus the same iron elements is presented in the panel (b) of the same figure. The two contours in the panel (a) are also plotted in this panel for comparison purpose. One can see that the overlap of the two contours and the distribution of the alpha elements in panel (b) is limited with [$\alpha$/Fe]$ <0.35$ dex. This is due to the limitation of the alpha elements in our calibration, i.e. $-0.05\leq[\alpha$/Fe$]\leq 0.35$ dex, while the spectroscopic alpha elements in SEGUE tend up to [$\alpha$/Fe]$\leq 0.5$ dex. Hence, the limited overlapping in question is a result of  a limited range of colour-[$\alpha$/Fe] calibration but not statistical.

\begin{figure}[h]
\begin{center}
\includegraphics[trim=0cm 0cm 1cm 0cm, clip=true, scale=0.42]{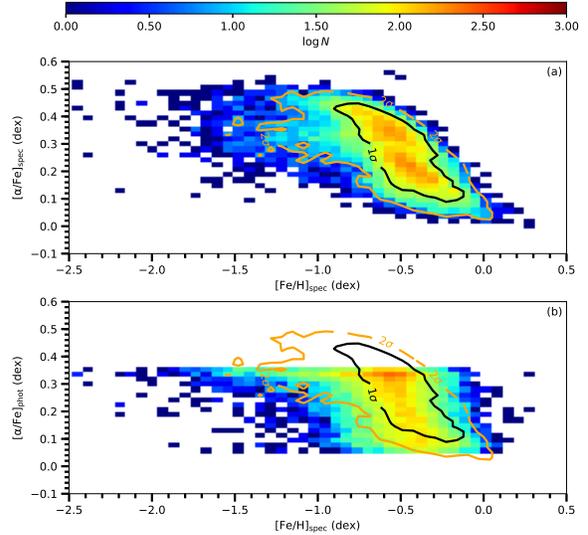}
\caption[]{Comparison of the spectroscopic alpha elements (panel a) with the photometric ones (panel b). The two contours cover the number of stars within one and two standard deviations of the total stars ($N=13,379$) taken from SEGUE.} 
\end{center}
\end {figure}

\section{Results}
The results obtained from the application of the calibration of the [$\alpha$/Fe] elements in terms of the $UV$-excess ($\delta$) defined in the $ugr$ system showed that this  calibration can be used in the determination of the alpha elements of the dwarf stars. Thus, we have two calibrations which can be used to provide alpha elements photometrically, i.e. one defined with the $UBV$ data in K16 and the next one defined with the $ugr$ data in this study. The first calibration has the advantage of covering more accurate $UBV$ data of the nearby stars, while the second one which is based on the $ugr$ data transformed from the $UBV$ ones used in the first calibration, has the advantage of its application to the stars observed in SDSS. We compared the results obtained from the two calibrations by means of the procedure explained in the following and noticed that there is a good agreement between them. Thus, we confirmed the reliability of the calibration defined with the $ugr$ data which are available for a large survey, SDSS, by comparison its results with the ones defined with the $UBV$ data which are observed for the nearby stars and that they are more accurate.   

We transformed the $(g-r)_0$ and $(u-g)_0$ colour-indices of the dwarf sample, mentioned in Section 3.2.1, to the $(B-V)_0$ and $(U-B)_0$ ones using the inverse transformation equations and estimated their [$\alpha$/Fe] elements via the procedure  in K16. Then, we combined these elements with their vertical distances relative to the Galactic plane, $z$, and plotted them in Fig. 7. The trend of the distribution consists of two segments as in Fig. 4, i.e. one with a small positive gradient, $d[\alpha/{\rm Fe}]/dz=+0.041\pm0.002$ dex kpc$^{-1}$ up to 5 kpc, and another one which is almost flat ($d[\alpha/{\rm Fe}]/dz=-0.007\pm0.002$ dex kpc$^{-1}$).	             

\begin{figure}[t]
\begin{center}
\includegraphics[scale=0.27, angle=0]{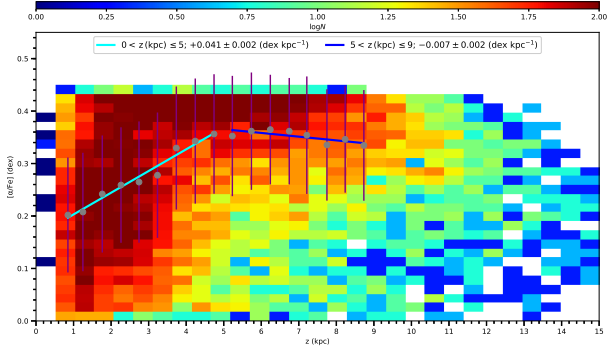}
\caption{Same as in Figure 4, but based on $UBV$ data using K16 calibration.} 
\end{center}
\end {figure}

A second confirmation of our argument that the alpha elements can be determined photometrically is carried out by comparison of the photometric alpha elements with the synthetic ones as explained in the following. We estimated [$\alpha/{\rm Fe}]_{syn}$, synthetic alpha-elements, and $\delta_{syn}$, synthetic {\it UV}-excesses, for our sample, $0.1<(g-r)_0 \leq 0.6$ mag, using Dartmouth Stellar Evolution Program \citep[DSEP,][]{Dotter08} and compared their distributions with the ones estimated by our calibration. We took the DSEP isochrones and did the necessary interpolations to get the required data for the relations. The estimations of $[\alpha/{\rm Fe}]_{syn}$ and $\delta_{syn}$ are carried out for three populations, thin disc, thick disc and halo whose metallicity ranges are assumed to be $-0.5<{\rm[Fe/H]}\leq +0.5$, $-1.0<{\rm[Fe/H]}\leq -0.5$, and $-2.5<{\rm[Fe/H]} \leq -1.0$ dex, respectively. Also, we adopted the 3, 12 and 13 Gyr as the ages of the populations in the same order. Thus, we estimated a set of ($g-r$, $u-g$) couples for each population using an iron abundance, an [$\alpha$/Fe]$_{syn}$ value and an age value, each time. The range of the $[\alpha/{\rm Fe}]_{syn}$ is adopted as $-0.2<[\alpha/{\rm Fe}]_{syn}\leq +0.8$. The iron metallicity and [$\alpha$/Fe]$_{syn}$ are used in 0.5 and 0.2 dex steps, respectively. Finally, we considered the $g-r$ and $u-g$ colour indices which lie in the range $0.1 <(g-r) \leq 0.6$ mag. The {\it UV}-excesses are estimated relative to the {\it UV} index for [Fe/H]=0 dex. We could fit the $[\alpha/{\rm Fe}]_{syn}$ and $\delta_{syn}$ data to a third degree polynomial with a high correlation coefficient ($R^2=0.99$): [$\alpha$/Fe]$_{syn}= 0.017\times\delta^3+0.264\times\delta^2+10.603\times \delta-22.826$, and compared it with the observational one obtained in our study. Fig. 8 shows that the two calibrations have the same trends. Also, for $\delta_{syn}<0.15$ mag the synthetic curve lies within the region occupied by the calibration curve. Although the synthetic curve for larger $\delta_{syn}$ occupies the alpha elements which are a bit larger than the ones corresponding to the calibration curve, an additional error of $\Delta [\alpha/{\rm Fe}]=0.1$ dex to our calibration supplies agreement also for this segment. The agreement of the synthetic calibration with the one based on the photometric alpha elements indicates that the alpha abundances of individual stars can be determined via {\it UV}-excess.

\begin{figure}[h]
\begin{center}
\includegraphics[scale=0.55, angle=0]{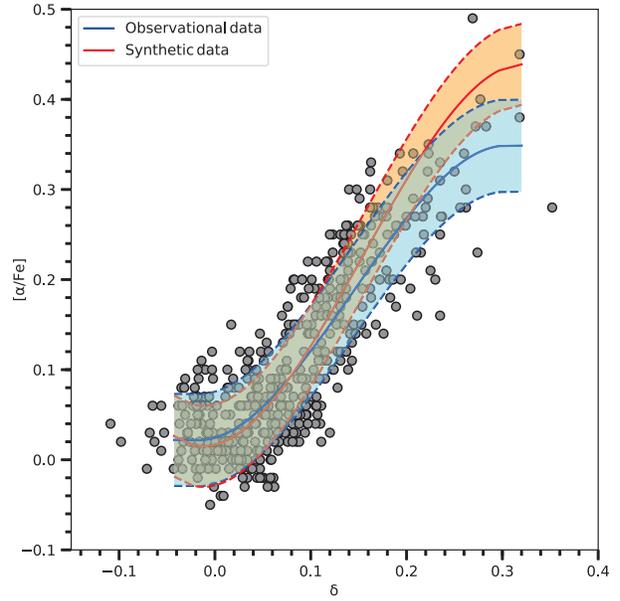}
\caption[]{Colour-coded comparison of the observational calibration with the synthetic one within an error of $\Delta[\alpha/{\rm Fe}]=0.1$ dex. Full circles correspond to the individual stars used in the observational calibration.} 
\end{center}
\end {figure}

\section{Summary and Discussion}
We calibrated the [$\alpha/$Fe] elements of 465 dwarf stars, taken from V04, B14, R06 and N10, in terms of the $UV$-excess $\delta$ defined in the $ugr$ system. The $(g-r)_0$ and $(u-g)_0$ data used in our evaluations are transformed from the $(B-V)_0$ and $(U-B)_0$ data in K16. The ranges of the $(g-r)_0$, $(u-g)_0$, $\delta$, and [$\alpha/$Fe] are $0.1<(g-r)_0<0.6$, $0.85<(u-g)_0<1.71$, $-0.11<\delta<0.30$ mag and $-0.05 < [\alpha/{\rm Fe}]<0.35$ dex, respectively. We used a single $[\alpha/{\rm Fe}]\times \delta$ calibration for all stars because the $(g-r)_0$ range is small and majority of the sample stars occupy the bluer part of the two-colour diagram (Fig. 1) where the guillotine effect is small. The squared of the correlation coefficient and the standard deviation of the unique calibration are at the level of nine calibrations carried out in K16, i.e. $R^2=0.74$ and $\sigma=0.05$ dex.
 
We applied the procedure to a high-latitude SDSS field, $85^\circ \leq b \leq 90^\circ$ with size 78 deg$^2$ which covers 23,414 F and G dwarfs. The $(u-g)_0$ and $(g-r)_0$ colour indices of these stars, taken from \cite{Guctekin17}, are used to estimate their $UV$-excess, $\delta$, which provide [$\alpha/$Fe] elements via our calibration, i.e. Eq. (3). Our star sample covers a large vertical distance interval relative to the Galactic plane, $0<z<9$ kpc (Fig. 4). The trend of [$\alpha/$Fe] consists of two segments: the first one has a positive gradient, $d[\alpha/{\rm Fe}]/dz=+0.032 \pm 0.002$ dex kpc$^{-1}$, and can be defined with $0<z<5$ kpc and $0.1<[\alpha/{\rm Fe}]<0.3$ dex which indicates a smooth transition from thin disc stars to the thick disc/halo stars. While the second segment covers the larger z distances and has almost a flat distribution, i.e. $[\alpha/{\rm Fe}]\approx 0.30$ dex. 

Our calibration with ranges $-0.1<\delta \leq 0.3$ mag and $-0.05<[\alpha/{\rm Fe}]\leq0.35$ dex could not be used for transformation of the $UV$-excesses larger than $\delta=0.30$ mag to the [$\alpha/$Fe] elements. Most of the alpha elements given in the literature \citep[cf. Hypatia catalogue,][and references therein]{Hinkel14} are limited with $[\alpha/{\rm Fe}]<0.4$ dex. Hence, the current calibration is limited to $[\alpha/{\rm Fe}]<0.4$ dex, and is not applicable for high-alpha abundance stars, such as the ones observed in the $Gaia$-ESO Survey \citep{Gilmore12}. The upper limit of the alpha elements measured by \cite{Jackson14} is $[\alpha/{\rm Fe}]=0.6$ dex. These authors used 381 metal-poor ([Fe/H]$<-1$ dex) halo stars and separated them into two categories, low-$\alpha$ and high-$\alpha$, which are defined with $0<[\alpha/{\rm Fe}]\leq 0.3$ and $0.3<[\alpha/{\rm Fe}] \leq 0.6$ dex, respectively. It seems that our sample consists of low-$\alpha$ abundances. 

\section*{Acknowledgments} 
Authors are grateful to the anonymous referee for his/her considerable contributions to improve the paper. This research has made use of NASA (National Aeronautics and Space Administration)'s Astrophysics Data System and the SIMBAD and VizieR Astronomical Database, operated at CDS, Strasbourg, France.


\begin{thebibliography}{}

\bibitem[Ahn et al.(2014)]{Ahn14}
Ahn, C. P., Alexandroff, R., Allende Prieto, C., et al., 2014, ApJS, 211, 17 

\bibitem[Alam et al.(2015)]{Alam15} 
Alam, S., et al., 2015, ApJS, 219, 12

\bibitem[Bensby et al.(2014)]{Bensby14}
Bensby, T., Feltzing, S., Oey, M.~S., 2014, A\&A, 562, A71

\bibitem[Bilir et al.(2008)]{Bilir08} 
Bilir, S., Ak, S., Karaali, S., Cabrera-Lavers, A., Chonis, T. S., Gaskell, C. M., 2008, MNRAS, 384, 1178

\bibitem[Caffau et al.(2011)]{Caffau11}
Caffau, E., Bonifacio, P., Francois, P., et al., 2011, Natur, 477, 67 

\bibitem[Carney(1979)]{Carney79}
Carney, B.~W., 1979, AJ, 84, 515 

\bibitem[Chonis \& Gaskell(2008)]{Chonis08}
Chonis, T.~S., Gaskell, C.~M., 2008, AJ, 135, 264 	

\bibitem[Dotter et al.(2008)]{Dotter08}
Dotter, A., Chaboyer, B., Jevremovic, D., Kostov, V., Baron, E., Ferguson, J. W., 2008, ApJS, 178, 89

\bibitem[Ecuvillon et al.(2006)]{Ecuvillon06} 
Ecuvillon A., Israelian G., Santos N.~C., Mayor M., Gilli G., 2006, A\&A, 449, 809 

\bibitem[Galeev et al.(2004)]{Galeev04}
Galeev, A.~I., Bikmaev, I.~F., Musaev, F.~A., Galazutdinov, G.~A., 2004, ARep, 48, 492

\bibitem[Gilmore \& Reid(1983)]{Gilmore83}
Gilmore, G., Reid, N., 1983, MNRAS, 202, 1025	

\bibitem[Gilmore et al.(2012)]{Gilmore12}
Gilmore, G., Randich, S., Asplund, M., et al., 2012, Msngr, 147, 25
 
\bibitem[Heijmans et al.(2012)]{Heijmans12} 
Heijmans, J., Asplund, M., Barden, S., et al., 2012, Ground-based and Airborne Instrumentation for Astronomy IV. Proceedings of the SPIE, Volume 8446, article id. 84460W, 17 pp 

\bibitem[Helmi et al.(2003)]{Helmi03} 
Helmi, A., et al., 2003, ApJ, 586, 195

\bibitem[Hinkel et al.(2014)]{Hinkel14}
Hinkel, N.~R., Timmes, F.~X., Young, P.~A., Pagano, M.~D., Turnbull, M.~C., 2014, AJ, 148, 54

\bibitem[Jackson-Jones et al.(2014)]{Jackson14}
Jackson-Jones, R., Jofre, P., Hawkins, K., et al., 2014, A\&A, 571L, 5

\bibitem[Juri{\'c} et al.(2008)]{Juric08} 
Juri{\'c}, M., Ivezic, Z., Brooks, A., et al., 2008, ApJ, 673, 864

\bibitem[Karaali et al.(2003a)]{Karaali03a}
Karaali, S., Bilir, S., Karata\c s, Y., Ak, S. G., 2003a, PASA, 20, 165

\bibitem[Karaali et al.(2003b)]{Karaali03b}
Karaali, S., Ak, S. G., Bilir, S., Karata\c s, Y., Gilmore, G., 2003b, MNRAS, 343, 1013

\bibitem[Karaali et al.(2003c)]{Karaali03c}
Karaali, S., Karata\c s, Y., Bilir, S., Ak, S. G., Hamzao{\u g}lu, E., 2003c, PASA, 20, 270

\bibitem[Karaali et al.(2005)]{Karaali05}
Karaali, S., Bilir, S., Tun{\c c}el, S., 2005, PASA, 22, 24
 
\bibitem[Karaali et al.(2011)]{Karaali11} 
Karaali, S., Bilir, S., Ak, S., Yaz, E., Co\c skuno\u glu, B., 2011, PASA, 28, 95

\bibitem[Karaali et al.(2016)]{Karaali16} 
Karaali, S., Yaz G\"ok\c ce, E., Bilir, S., 2016, Ap\&SS, 361, 354

\bibitem[Minchev(2017)]{Minchev17}
Minchev, I., 2017, arXiv:1701.07034 

\bibitem[Neuforge-Verheecke \& Magain(1997)]{Neuforge97}
Neuforge-Verheecke, C., Magain, P., 1997, A\&A, 328, 261

\bibitem[Nissen \& Schuster(2010)]{Nissen10}
Nissen, P.~E., Schuster, W.~J., 2010, A\&A, 511L, 10

\bibitem[Petigura \& Marcy(2011)]{Petigura11}
Petigura, E.~A., Marcy, G.~W., 2011, ApJ, 735, 41	

\bibitem[Porto de Mello et al.(2008)]{Porto08} 
Porto de Mello, G.~F., Lyra, W., Keller, G.~R.,  2008, A\&A, 488, 653 

\bibitem[Reddy et al.(2006)]{Reddy06}
Reddy, B.~E., Lambert, D.~L., Allende Prieto, C., 2006, MNRAS, 367, 1329

\bibitem[Roman(1955)]{Roman55}
Roman, N.~G., 1955, ApJS, 2, 195

\bibitem[Sandage(1969)]{Sandage69}
Sandage, A., 1969, ApJ, 158, 1115

\bibitem[Steinmetz et al.(2006)]{Steinmetz06}
Steinmetz M., et al., 2006, AJ, 132, 1645

\bibitem[Stephens \& Boesgaard(2002)]{Stephens02}
Stephens, A., Boesgaard, A.~M., 2002, AJ, 123, 1647

\bibitem[Tun{\c c}el G{\"u}{\c c}tekin et al.(2016)]{Guctekin16}
Tun{\c c}el G{\"u}{\c c}tekin, S., Bilir, S., Karaali, S., Ak, S., Ak, T., Bostanc{\i}, Z.~F., 2016, Ap\&SS, 361, 186 

\bibitem[Tun{\c c}el G{\"u}{\c c}tekin et al.(2017)]{Guctekin17}
Tun{\c c}el G{\"u}{\c c}tekin, S., Bilir, S., Karaali, S., Plevne, O., Ak, S., Ak, T., Bostanc{\i}, Z.~F., 2017, Ap\&SS, 362, 17 

\bibitem[Valenti \& Fischer(2005)]{Valenti05}
Valenti, J.~A., Fischer, D.~A., 2005, ApJS, 159, 141

\bibitem[Venn et al.(2004)]{Venn04}
Venn, K.~A., Irwin, M., Shetrone, M.~D., et al., 2004, AJ, 128, 1177

\bibitem [Wildey et al.(1962)]{Wildey62}
Wildey, R.~L., Burbidge, E.~M., Sandage, A.~R., Burbidge, G.~R., 1962, ApJ, 135, 94

\bibitem [Wilson et al.(2010)]{Wilson10}
Wilson, J. C., Hearty, F., Skrutskie, M. F., et al., 2010, Proceedings of the SPIE, Volume 7735, id.

\bibitem[Wyse \& Gilmore(1988)]{Wyse88}
Wyse, R.~F.~G., Gilmore, G., 1988, AJ, 95, 1404

\bibitem[Yanny et al.(2009)]{Yanny09}
Yanny, B., Rockosi, C., Newberg, H. J., et al., 2009, AJ, 137, 4377

\bibitem[Zoccali et al.(2014)]{Zoccali14}
Zoccali, M., Gonzalez, O. A., Vasquez, S., et al., 2014, A\&A, 562A, 66 

\end{thebibliography}
\end{document}